%% file: Manuscript.tex
\documentclass[a4paper, amsfonts, amssymb, amsmath, reprint, showkeys, nofootinbib, twoside]{revtex4-1}
\usepackage[english]{babel}
\usepackage[utf8]{inputenc}
\usepackage[colorinlistoftodos, color=green!40, prependcaption]{todonotes}
\input{preamble}
\usepackage[pdftex, pdftitle={Article}, pdfauthor={Author}]{hyperref} 
 \usepackage[normalem]{ulem}
  \usepackage{soul}
  \usepackage{graphicx}
 \usepackage{xcolor} 
\begin{document}
\title{Feedback-Driven Ground-State Search in Coupled Laser Arrays}

\author{Rajneesh Fulara$^{1}$}
\author{Fabien Bretenaker$^{2}$}
\author{Vishwa Pal$^{1}$}

\email[Correspondence email address:]{vishwa.pal@iitrpr.ac.in}

\affiliation{$^{1}$Department of Physics, Indian Institute of Technology Ropar, Rupnagar 140001, Punjab, India.}
\affiliation{$^{2}$Université Paris-Saclay, ENS Paris-Saclay, CNRS, CentraleSupélec, LuMIn, Orsay, France.}
\date{\today} 

\begin{abstract}
Optimisation problems, which appear in numerous fields of science and industry, are challenging to solve even with modern supercomputers. Alternatively, many of these problems can be mapped onto the ground-state search of the spin Hamiltonian, which has been physically implemented on various platforms whose intrinsic dynamics are analogous to spin systems, such as optical parametric oscillators, Bose-Einstein condensates, and laser arrays. However, the complex energy landscape of spin Hamiltonians often traps the system in local minima, preventing the system from reaching the ground-state (global minimum). Here, we demonstrate an intrinsic, feedback-driven annealing mechanism in class-B semiconductor laser arrays, arising from the interplay between the internal coupling $\alpha$ and the external coupling $\eta$, in which the instantaneous phase configuration of the lasers self-modulates their amplitude fluctuations. These amplitude fluctuations act as temperature, dynamically reshaping the effective potential and enabling the system to escape local minima. Using a one-dimensional (1D) ring array of lasers, we analyzed defect formation in the $\alpha$–$\eta$ parameter space and identified an optimal regime in which the system reaches the ground-state with $\sim100\%$ probability. Although both $\alpha$ and $\eta$ are required for the feedback loop, it helps to suppress defect formation by modifying two competing timescales, amplitude stabilization time ($t_\mathrm{amp}$) and phase locking time ($t_\mathrm{phase}$), in analogy with the Kibble–Zurek (KZ) mechanism. These timescales can be varied independently by changing either $\alpha$ or $\eta$ while keeping the other fixed.  When feedback produces the same time scale ratio, it yields the same defect probability, irrespective of whether the time scales are tuned via $\alpha$ or $\eta$. This universality confirms that the relative time scales, not specific parameter, determines the formation of the defect in laser arrays. Our results establish internal feedback-driven annealing as a practical route to ground-state search in semiconductor laser arrays, providing a foundation for efficient, scalable laser-based spin simulators capable of tackling hard optimization problems.
\end{abstract}
\maketitle

\section{Introduction}
Combinatorial optimization problems, which appear across many domains including finance, drug discovery, artificial intelligence and social networks, are difficult to solve even with advanced algorithms and modern supercomputers because of their NP-hard (nondeterministic polynomial-time ) nature \cite{kitchen2004,lucas2014,tanahashi2019,hauke2020,hopfield1982,shniderman2024}. A broad class of these problems can be mapped to ground-state search problem of the spin Hamiltonian, such as the Ising and XY models \cite{lucas2014,berloff2017}. Therefore, there is a significant growing interest in physical platforms that emulate spin Hamiltonians, which enables the finding of solutions to optimisation problems by experimentally observing the ground-state. These spin simulators have been realized in different systems, including superconducting circuits \cite{boixo2014,king2018}, Bose-Einstein-condensate polaritons \cite{kalinin2020,berloff2017,kalinin2018}, stochastic magnetic junctions \cite{borders2019}, memristors \cite{cai2020}, coupled electrical oscillators \cite{shukla2014}, complementary metal oxide semiconductor devices (CMOS) \cite{merolla2014}, time-multiplexed optical parametric oscillators (OPOs) \cite{mcmahon2016,inagaki2016,inagaki2016-1,marandi2014}, photonic circuits \cite{roques2020,prabhu2020,shen2017,okawachi2020}, spatial light modulators \cite{pierangeli2019,pierangeli2020,pierangeli2020-1,veraldi2025,sakellariou2025,yamashita2023}, and coupled laser arrays \cite{nixon2013,gershenzon2020,utsunomiya2011,takata2012,babaeian2019,tradonsky2019}. 

Compared to other physical platforms, optical platforms are becoming particularly interesting because they offer better scalability, all-to-all connectivity between spins, inherent parallelism, low power consumption, and room-temperature operation. Among these, coupled laser arrays are especially promising because their intrinsic nonlinear dynamics allow spin interactions and energy minimization to be performed entirely within the optical domain, eliminating the need for external electronic control, as in OPOs and SLM-based Ising machines. In coupled lasers, spin is represented by the laser's phase, while the total loss of the system corresponds to the energy of an effective spin Hamiltonian \cite{nixon2013,gershenzon2020}. Being gain-dissipative systems, the lasers naturally seek to minimize their total loss, which is equivalent to finding the ground-state of the spin system \cite{nixon2013,honari2020}. Networks with thousands of spins have been realized in coupled laser arrays \cite{nixon2013}. These systems have demonstrated their versatility by simulating spin systems with artificial gauge fields \cite{mahler2025}, sampling the ground-state of the XY Hamiltonian \cite{pal2020}, and solving phase-retrieval problems \cite{tradonsky2019}, highlighting their potential as a powerful spin simulator.

\begin{figure}[htbp]
\centering
\includegraphics[height =7.9cm, keepaspectratio = true]{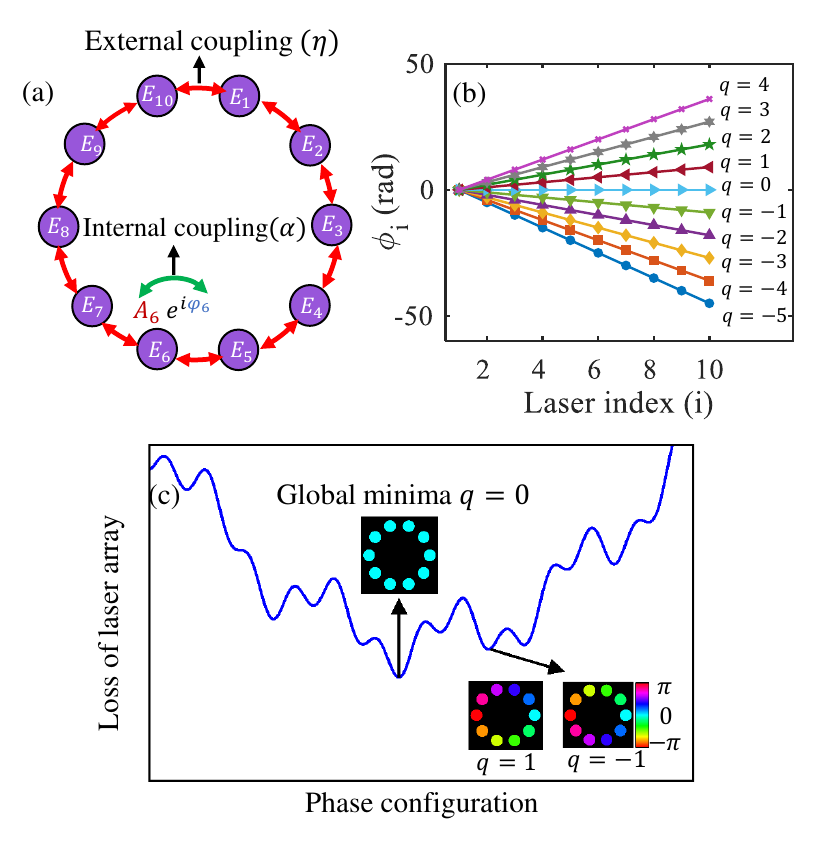}
\caption{(a) One-dimensional ring array of $N=10$ semiconductor lasers with dissipative coupling $\eta$ between nearest neighbours (external coupling). The factor $\alpha$ provides amplitude-phase coupling within each laser (internal coupling). (b) The corresponding steady-state phase-locked solutions are $\phi_i = i q 2\pi/10$, where $i$ is the laser index and $q$ is the topological charge. Here, $q=0$ represents the in-phase state and $q\neq0$ denotes topological defect states. (c) The schematic loss landscape, where the in-phase state ($q=0$) is the global minimum, while states with topological defects ($q\neq0$) correspond to local minima.}
\label{Fig_1}
\end{figure}

However, as the phases evolve over a complex loss landscape with many local minima, the lasers can become stuck in a local minimum instead of the global minimum, posing a major challenge for their use as optimization solvers \cite{pal2017, karuseichyk2025,bouchereau2022}. In positively coupled 1D ring laser arrays, the in-phase steady-state, where all lasers have the same phase, corresponds to the global minimum, and states with non-zero phase differences, also known as topological defect states, represent local minima in the loss landscape (Fig.\,\ref{Fig_1}). In spin systems, annealing (heating and slow cooling) prevents the spins from becoming trapped in local minima (topological defects) via thermal fluctuations. This principle of using controlled fluctuations has been applied to optical systems to overcome topological defects. For instance, in coupled lasers and oscillators, injected disorder or noise has been demonstrated to help avoid topological defects and improve phase locking and synchronization \cite{prabhu2020,pierangeli2020,pando2024,nair2021}.  Despite their effectiveness, these methods increase system complexity by requiring the precise external injection of controlled disorder.

In contrast, an intrinsic annealing mechanism exists in class-B lasers. The amplitude fluctuations, due to initial relaxation oscillations, are initially high and gradually decay to zero. These fluctuations act as an effective temperature, providing an intrinsic annealing mechanism that helps to reduce topological defects \cite{pal2017}. Similar to annealed spin systems, the probability of defect formation in coupled laser arrays is related to the Kibble–Zurek (KZ) mechanism, where the probability of defects depends on the competition between the quench rate and the system's relaxation time \cite{kibble1976topology,zurek1985cosmological, pal2017}.In practice, however, annealing requires infinitely slow cooling to fully suppress defects \cite{kirkpatrick1983}. In laser systems, this corresponds to operating very close to threshold to achieve an infinitely slow decay of amplitude fluctuations, which is very difficult to implement experimentally; consequently, defects are always present.

Recent studies have highlighted the linewidth enhancement factor $\alpha$ as a key parameter for suppressing topological defects in coupled semiconductor lasers \cite{bouchereau2022}. In these arrays, two distinct couplings govern the dynamics (Fig. \ref{Fig_1}): an internal coupling $\alpha$, which links the amplitude and phase within a single laser, and an external coupling $\eta$, which couples the optical fields between different lasers. While the beneficial effect of $\alpha$ is observed, the underlying physical mechanism through which it modifies the annealing process remains entirely unexplored within the spin-simulator framework. Moreover, these studies have been limited to specific values of $\eta$ and $\alpha$. Therefore, a systematic investigation across the $\alpha$–$\eta$ parameter space is required to determine whether the beneficial effects of $\alpha$ in defect suppression are general or confined to specific parameter regimes of $\eta$, and to elucidate how it modifies the effective annealing process in class-B lasers to guide the system toward the ground-state. Understanding the behavior of $\alpha$ across the full parameter space of $\alpha$, and $\eta$ is essential to establishing the conditions under which laser arrays can be reliably transformed into controllable optimization devices.

In this work, we numerically investigate topological defect formation across the $\alpha$–$\eta$ parameter space in a one-dimensional ring array of class-B semiconductor lasers. We demonstrate that within a specific region of this space, $\alpha$ completely suppresses topological defects and guides the system to its ground state. The defect suppression results from an intrinsic nonlinear feedback loop generated by the interplay between the internal ($\alpha$) and external ($\eta$) couplings. This feedback loop causes amplitude fluctuations, which act as an effective temperature, to dynamically increase or decrease depending on the instantaneous phase configuration. We term this process intrinsic feedback–driven adaptive annealing, where the system self-regulates its fluctuations (effective temperature) based on its phase state. This intrinsic feedback dynamically reshapes the effective potential governing the laser dynamics, thereby preventing the system from becoming trapped in local minima. Crucially, while both parameters $\alpha$ and $\eta$ are required to establish the feedback loop, we find that defect suppression depends on two intrinsic timescales: the amplitude stabilization time ($t_{\text{amp}}$) and the phase-locking time ($t_{\text{phase}}$). By analyzing $t_{\text{amp}}$ and $t_{\text{phase}}$ across the parameter space, we show that the defect probability depends solely on their ratio $t_{\text{amp}}/t_{\text{phase}}$, collapsing all data onto a single universal curve, similar to the KZ mechanism. This universal data collapse implies that any parameter pair $(\alpha, \eta)$ yielding the same timescale ratio produces identical annealing outcomes. Thus, defect suppression is governed by $t_{\text{amp}}/t_{\text{phase}}$ , with $\alpha$ and $\eta$ acting as independent knobs that can tune this ratio.

\section{Theoretical Model}
To investigate the intrinsic feedback-driven annealing mechanism in semiconductor laser arrays for efficiently finding the ground-state, we consider a uniform nearest-neighbour-coupled 1D ring array of lasers (Fig.\,\ref{Fig_1}(a)). The dynamics of a single-transverse, single-longitudinal mode laser in this array is governed by the following rate equations, with time normalized to the cavity roundtrip time, $\tau_{\text{cav}}$ \cite{fabiny1993,rogister2004,bouchereau2022}:
\begin{eqnarray}
  \frac{dA_i}{dt} &=& -\frac{1}{2} \left(1 - \frac{n_i}{n_{\text{th}}} \right) A_i + \frac{\eta}{2}\Big[ \cos(\phi_{i+1} - \phi_i) \nonumber \\
  && A_{i+1} + \cos(\phi_{i-1} - \phi_i) A_{i-1} \Big], \label{Eq1}\\
 \frac{d\phi_i}{dt}&=& \frac{\alpha}{2} \frac{n_i}{n_{\text{th}}} + \frac{\eta}{2} \Big[ \frac{A_{i+1}}{A_i} \sin(\phi_{i+1} - \phi_i) \nonumber \\
    &+& \frac{A_{i-1}}{A_i} \sin(\phi_{i-1} - \phi_i) \Big] + \omega_i, \label{Eq2}\\
 \frac{dn_i}{dt} &=&  \frac{1}{T}\left(r_i n_{\text{th}} - {n_i} -\frac{A_i^2}{F_{\text{sat}}} n_i\right). \label{Eq3}
\end{eqnarray}
Here, $A_i$, $\phi_i$, and $n_i$ denote the amplitude, phase, and carrier density of the $i_\mathrm{th}$ laser in a 1D ring array, respectively, where the index $i$ labels the individual lasers. Throughout this work, time will be expressed in units of the cavity lifetime $\tau_{\text{cav}}$. The $\eta$ represents the dissipative coupling between neighboring lasers, which we denote as external coupling. In contrast, $\alpha$ provides an internal coupling between the amplitude and phase of a single laser, mediated by the carrier density $n$, as shown in Fig.\,\ref{Fig_1}(a) \cite{henry1982}. Its value is significant in semiconductor lasers ($\alpha\ \approx 2-10$ ), but negligible in solid state and gas lasers (less than 1) \cite{zilkie2008,sinquin2023,consoli2012,thorette2017}. Due to $\alpha$, semiconductor lasers exhibit unique properties like broader spectral linewidth and chaotic dynamics, and are inherently more sensitive to noise and feedback than solid-state lasers \cite{wang1988,winful1988,winful1992,nair2018,winful1990}. The site-dependent excitation ratio or pumping rate and frequency detuning are given by $r_i$ and $\omega_i$, respectively; however, we restrict our analysis to the homogeneous case $r_i = r$ and $\omega_i = 0$ for all $i$. The parameters $n_{\mathrm{th}}$ and $F_{\mathrm{sat}}$ denote threshold carrier density, and saturation photon number, respectively and are identical for all lasers. The dimensionless parameter $T$ represents the ratio of carrier lifetime to cavity round-trip time $\tau_{cav}$. For our study, we use $T=1000$, corresponding to a standard class-B laser regime.  This class of lasers is chosen because it exhibits initial amplitude relaxation oscillations that act as an effective temperature, causing the laser array to behave as an annealed spin system \cite{winful1990,pal2017}. We numerically solved the Eqs.\,(\ref{Eq1})-(\ref{Eq3}) using MATLAB-ODE 45 solver for $1D$ ring array of $N=20$ lasers. This system size is chosen because it is computationally tractable and captures the essential physical mechanism. The initial amplitudes and phases are chosen randomly from uniform distributions over $[0,1]$ and $[-\pi,\pi]$, respectively, with the initial carrier density set to zero. The parameter $F_{\mathrm{sat}}$ sets the $n_{th}$ and acts as a pure scaling factor; it does not affect the qualitative dynamics of the system. For all simulations, we set $F_{\mathrm{sat}} = 10^{10}$.

At steady-state, a 1-D ring laser array of finite size $N$ possesses a finite number distinct phase-locked configurations \cite{pal2017}. These states are characterized by an adjacent-laser phase difference $\phi_{i+1}-\phi_i= \Delta\phi = q \frac{2\pi}{N}$, where $q$ is an integer ranging from $q = -\frac{N-1}{2}$ to $ \frac{N-1}{2}$ for odd $N$ ($q = -\frac{N}{2}$ to $\frac{N}{2}-1$ for even $N$). In a 1D ring array of $N=10$ lasers, the phase of each laser corresponding to different stable phase-locked states is shown in Fig.\,\ref{Fig_1}(b). The uniform in-phase state, $q=0$ is the minimum-loss state and corresponds to the global minimum of the loss landscape, whereas the states with non-zero phase difference, $q\neq0$ represents higher losses and correspond to local minima in the landscape (see Appendix\,\ref{sec:appendix A} and Appendix\,\ref{sec:appendix B}).

To identify the ground-state and topological defect states, we define the order parameter as
\begin{equation}
    \rho(t) = \left| \frac{1}{N} \sum_{i=1}^{N} e^{i\phi_i(t)} \right|,
\end{equation}
where $\phi_i(t)$ represents the phase of the $i_\mathrm{th}$ laser at time \(t\), and $N$ is the total number of lasers in the array. During evolution, if the array proceeds toward the ground-state ($q=0$ ), $\rho$ approaches $1$, whereas for topological defect states ($q\neq0$), it approaches 0 \cite{pal2017}. 
\begin{figure}[htbp]
\centering
\includegraphics[width=\columnwidth]{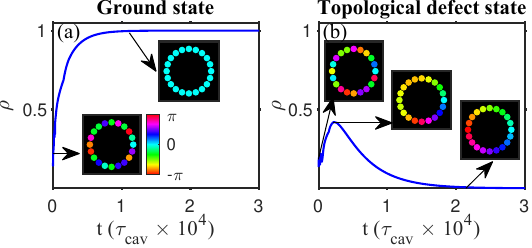}
\caption{In a 1-D ring array of $N=20$ lasers, time evolution of order parameter ($\rho$) leading to steady-state (a) ground state ($\rho=1$), and (b) topological defect state ($\rho=0$). The insets show the phase distributions at different times during evolution. The simulation parameters are $\eta=0.005$, $r=2$ and $\alpha$=0.}
\label{Fig_2}
\end{figure}

Figure\,\ref{Fig_2} shows the time evolution of $\rho$ for two different realizations corresponding to ground state and topological defect state. For each realization, the phases are initially random, and hence $\rho$ is close to $0$. As the system evolves toward the ground-state, $\rho$ increases and approaches $1$, as shown in Fig.\,\ref{Fig_2}(a). However, in the case of a topological defect state, although the system initially moves toward the ground-state, indicated by an increase in $\rho$, it becomes trapped in a local minimum and remains stable there, indicated by a decrease in $\rho$ over time, which approaches $0$ (Fig.\,\ref{Fig_2}(b)). The final state, whether it is a ground-state or a topological defect state, can be inferred by the value of $\rho$.

\section{The Feedback-Driven Effective Temperature Analogy}
In class-B lasers, the amplitude undergoes transient relaxation oscillations before stabilizing \cite{honari2021self}. These temporal amplitude variations can be quantified at any time $t$ using a time-dependent standard deviation, $\sigma(t)$, defined as:
\begin{equation}
\sigma(t) = \sqrt{\frac{1}{W}
\int_{t-\frac{W}{2}}^{t+\frac{W}{2}}
\left[ A(s) - \bar{A}(t) \right]^2 \, ds },
\end{equation}
where $A(s)$ is the amplitude at time $s$ and $\overline{A}(t)$ is the local mean computed over an time interval $W$ around $t$. A large value of $\sigma$ indicates significant amplitude fluctuations within the interval, while $\sigma \to 0$ represents that the amplitude has stabilized. We set $W = 20\thinspace \tau_{cav}$, an interval long enough to capture the amplitude variation yet short enough to resolve the slow decay of the relaxation oscillation envelope. Moreover, the results are insensitive to moderate variations of $W$.

\begin{figure}[htbp]
\centering
\includegraphics[width=\columnwidth]{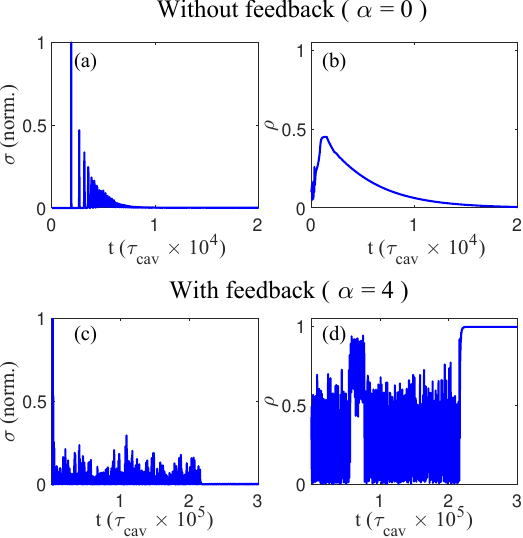}
\caption{Time evolution of $\sigma$ and $\rho$ in a 1-D ring array of 20 lasers (a, b) without feedback $\alpha=0$, and (c, d) with feedback $\alpha=4$. Simulation parameters are $\eta=0.005$ and $r=2$.}
\label{Fig_3}
\end{figure}

For $\alpha=0$, the amplitude fluctuation strength, quantified by $\sigma$, decays continuously over time, as indicated in Fig.\,\ref{Fig_3}(a). Since $\sigma$ acts as the system's effective temperature, this decay is analogous to the cooling process in thermal annealing of spin systems \cite{pal2017}. The cooling rate in lasers, i.e., how fast the laser is stabilising, can be controlled by the excitation ratio $r$ (see Appendix\,\ref{sec:appendix D}). This annealing helps to suppress topological defects similar to a spin system, as fluctuations aid the system in escaping local minima \cite{pal2017,inagaki2016-1,nixon2011}. A key limitation of the annealing mechanism is that it remains effective as long as the effective temperature remains high. At lower temperatures, thermal fluctuations are not strong enough to take the system out of local minima, and $\rho$ evolves smoothly toward a topological defect state (Fig.\,\ref{Fig_3}(b)) (\cite{rubin2017dual,zhang2019understanding}). Consequently, to suppress topological defects reliably, the cooling must be infinitely slow, requiring the pump to be tuned extremely close to threshold, a condition that is experimentally challenging to achieve. As a result, coupled laser arrays always have a finite probability of becoming trapped in local minima.

When $\alpha \neq 0$, the phase of each laser is directly influenced by its own amplitude via the carrier density $n_i$, through the term $\alpha n_i / n_{\text{th}}$ (Eq.\,(\ref{Eq2})). This establishes a bidirectional coupling: while the phases are affected by amplitudes, the amplitude dynamics simultaneously depend on the instantaneous phase configuration through the external coupling $\eta$. As a result, neither phase nor amplitude evolution is smooth. The phase evolution shows fluctuations in $\rho$ (Fig.\,\ref{Fig_3}(d)), and the amplitude evolution also becomes irregular (Fig.\,\ref{Fig_3}(c)), in contrast to the monotonic decay observed for $\alpha=0$ (Fig.\,\ref{Fig_3}(a)). 

This interplay between $\alpha$ and $\eta$ establishes a feedback loop. The external coupling $\eta$ controls how strongly phase differences generate amplitude fluctuations in neighbouring lasers, while internal coupling $\alpha$ governs how efficiently those fluctuations are converted back into phase changes. Consequently, the system's effective temperature, the amplitude fluctuation strength, $\sigma$, is no longer a simple decaying function. Instead, it is dynamically modulated by the phase configuration. This modulation is the core of an adaptive annealing mechanism. When the phase configuration is near a local minimum, the feedback loop increases $\sigma$, raising the effective temperature to provide the energy needed to escape. As the configuration approaches the global minimum, the feedback decreases $\sigma$, lowering the temperature to allow the system to settle into the stable ground-state, where both $\sigma$ and $\rho$ stabilize (Figs.\,\ref{Fig_3}(c) and \ref{Fig_3}(d)). This feedback-driven annealing mechanism therefore prevents the system from becoming permanently trapped in local minima and reliably guides it to its ground-state.

\section{Defect Probability and Phase Diagram}
\begin{figure*}[htbp]
\centering
\includegraphics[height = 13.5cm, keepaspectratio = true]{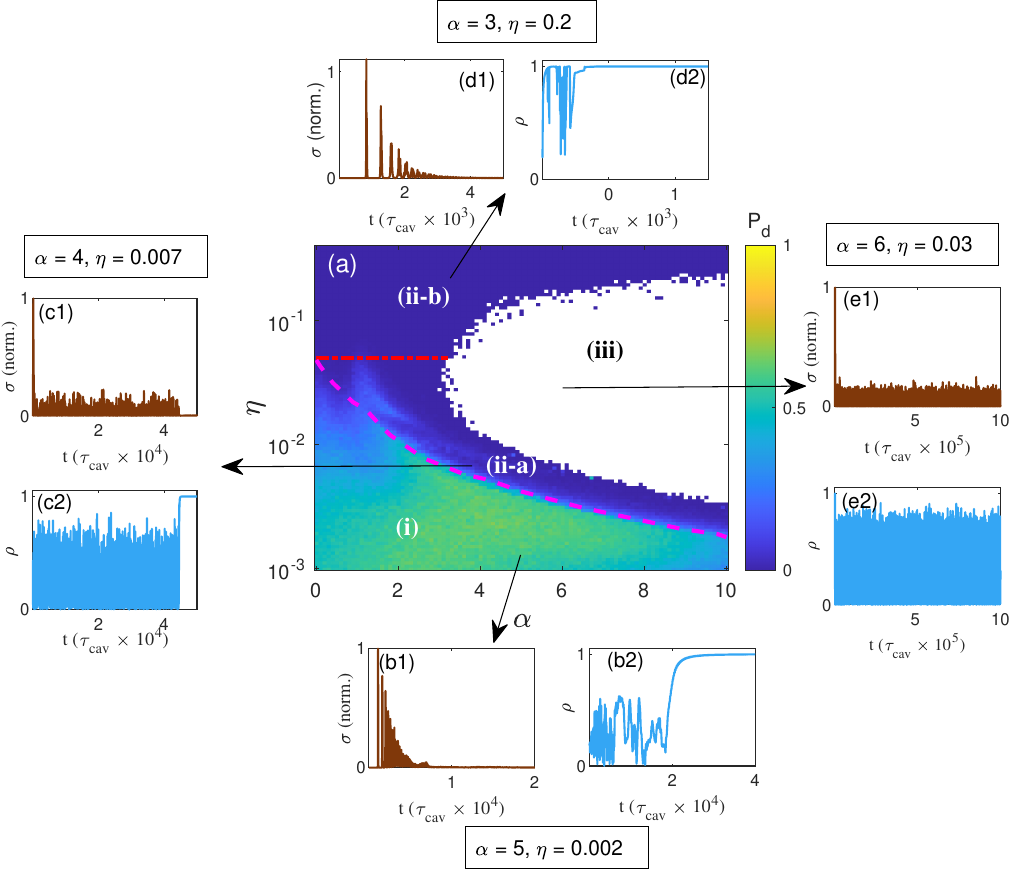}
\caption{In a 1-D ring array of $N=20$ lasers, the probability of topological defects ($\mathrm{P_d}$) as a function $\alpha$ and $\eta$ at a fixed value of $r=3$, indicating four different key regimes of laser array dynamics. Probability is determined by averaging the results over 2500 different random initial conditions. Note, in the white region the 1D ring array of lasers exhibits unstable behavior and hence no phase-locking is obtained. The insets indicate the representative dynamics of the order parameter $\rho$ and amplitude fluctuations $\sigma$ for each of the regions marked by the arrow.}
\label{Phase_plot}
\end{figure*}

To verify the effectiveness of feedback-driven annealing in reaching the ground state, we have analyzed the probability of topological defect formation ($\mathrm{P_{d}}$) across a broad parameter space of $\alpha$ and $\eta$. We numerically solved the laser rate equations (Eqs.\,(\ref{Eq1})-(\ref{Eq3})) for a 1D ring array of $N = 20$ lasers with an excitation ratio of $r = 3$, over the ranges, $0.001 \le \eta \le 0.4$ and $0 \le \alpha \le 10$. The results are shown in Fig.\,\ref{Phase_plot}, where the color represents $\mathrm{P_{d}}$ obtained by averaging over $2500$ random initial conditions. The dark blue region corresponds to a zero probability of topological defects ($\mathrm{P_{d}}=0$), while yellow represents a probability of one ($\mathrm{P_{d}}=1$); intermediate probabilities are indicated by the color gradient, as shown in the color bar. The white region corresponds to parameter sets that lead to unstable laser dynamics, preventing phase-locking.

From Fig.\,\ref{Phase_plot}(a), based on $\mathrm{P_{d}}$, we can identify three broad regimes: (i) a finite-defect region (colored gradient), (ii) a zero-defect region (dark blue), and (iii) an unstable region (white). Although the transitions between these regions are smooth, we have drawn visual guides to demarcate them for clarity. A pink dashed line separates the region with a finite probability of topological defects (region(i)) from the defect-free regions (region (ii)). The defect-free region can be further subdivided into two subregimes, (ii-a) and (ii-b), separated by a horizontal red dot-dashed line at $\eta = 0.05$, which distinguishes low-to-moderate coupling strengths in region (ii-a) from the very strong coupling regime (ii-b) with $\eta > 0.05$. The surrounding insets show the characteristic time evolution of the order parameter $\rho$ and the amplitude fluctuations $\sigma$ for a representative point in each regime. These figures illustrate how the interplay between $\alpha$ and $\eta$ shapes the transient dynamics that ultimately determine the formation of topological defects.

We now analyze each regime by examining the evolution of $\sigma$, which represents the effective temperature and thus captures the system's annealing behavior. As shown in the Fig.\,\ref{Phase_plot}(b1) for $(\alpha=5, \eta=0.002)$, in region (i), which lies below the pink dashed curve in the phase diagram, $\sigma$ decays monotonically from a high to a low value, similar to the case without feedback (Fig.\,\ref{Fig_3}(a)). This smooth decay indicates that the combined feedback from $\eta$ and $\alpha$ is insufficient to induce strong enough interaction between phase and amplitude dynamics, characteristic of normal annealing. Therefore, the system cannot explore phase space effectively, leading to a finite probability of defect formation Fig.\,\ref{Phase_plot}(b2). Region (ii-a) lies above the pink dashed curve, where we observe effective feedback-based annealing. In this regime, the interplay between $\eta$ and $\alpha$ is strong enough to create a dynamic coupling between phase and amplitude. As clearly seen in the Fig.\,\ref{Phase_plot}(c1) for $(\alpha=4, \eta=0.007)$, $\sigma$ no longer decays monotonically but exhibits fluctuations, increasing and decreasing, as the system actively explores different phase configurations and stabilizes only when a stable phase configuration is reached, corresponding to the ground state (Fig.\,\ref{Phase_plot}(c2)). This non-monotonic envelope of $\sigma$ is the signature of the feedback-based annealing process, which helps the system to escape local minima in the loss landscape. Consequently, this regime yields $\mathrm{P_{d}}\approx 0$, demonstrating the success of the feedback-based annealing mechanism.

While Region (ii-a) achieves zero defects through active feedback-based annealing, a second, physically distinct path to $P_{d}=0$ exists at very high external coupling. This defines Region (ii-b), the dark blue area above the horizontal red dot-dashed line at $\eta>0.05$. Here, the coupling is so strong that the phases of all lasers align almost instantaneously. Because the feedback mechanism modulates amplitude only in response to phase variations, this rapid phase locking suppresses any feedback-driven amplitude modulation. Consequently, $\sigma$ decays monotonically, similar to the case without effective feedback, as shown in Fig.\,\ref{Phase_plot}(d1) for ($\alpha=3,\eta=0.2$), and the laser array behaves as a single laser. This regime is of limited interest to us, as such extreme coupling is experimentally challenging to achieve (and becomes even more difficult for larger arrays).

Finally, Region (iii) corresponds to the white area in the phase diagram, where the combined feedback from $\alpha$ and $\eta$ drives the laser dynamics into an unstable regime, and no phase-locking occurs between the lasers. The dynamics are illustrated in Figs.\,\ref{Phase_plot}(e1)-\ref{Phase_plot}(e2) for ($\alpha=6,\eta=0.05$), where the $\sigma$ and $\rho$ continuously fluctuate with time without reaching a steady state. Such unstable behaviour in semiconductor lasers is well understood and consistent with previous studies \cite{winful1988, winful1990, winful1992}.

It is evident from the phase diagram that intrinsic feedback-based annealing is effective only when at least one of the two couplings, $\eta$ or $\alpha$, is strong enough to generate fluctuations capable of kicking the system out of local minima. The two parameters therefore act in complementary ways, as shown by the shape of the pink dashed line in Fig.\,\ref{Phase_plot}: a higher value of $\eta$ can compensate for a lower $\alpha$, and vice versa, to achieve effective annealing. However, excessive values of both parameters destabilise the lasers (Region iii), while values that are too small weaken the feedback loop altogether (Region i). This delicate balance defines a narrow region in the $(\alpha,\thinspace\eta)$ space where feedback-based annealing is most effective (Region ii-a). Here, the interplay between internal $\alpha$ and external coupling $\eta$ yields the strongest suppression of topological defects, demonstrating that the underlying mechanism depends on their combined effect rather than on either parameter alone.

\section{Description based on Effective Potential}
\begin{figure*}[t]
\centering
\includegraphics[height = 11cm, keepaspectratio = true]{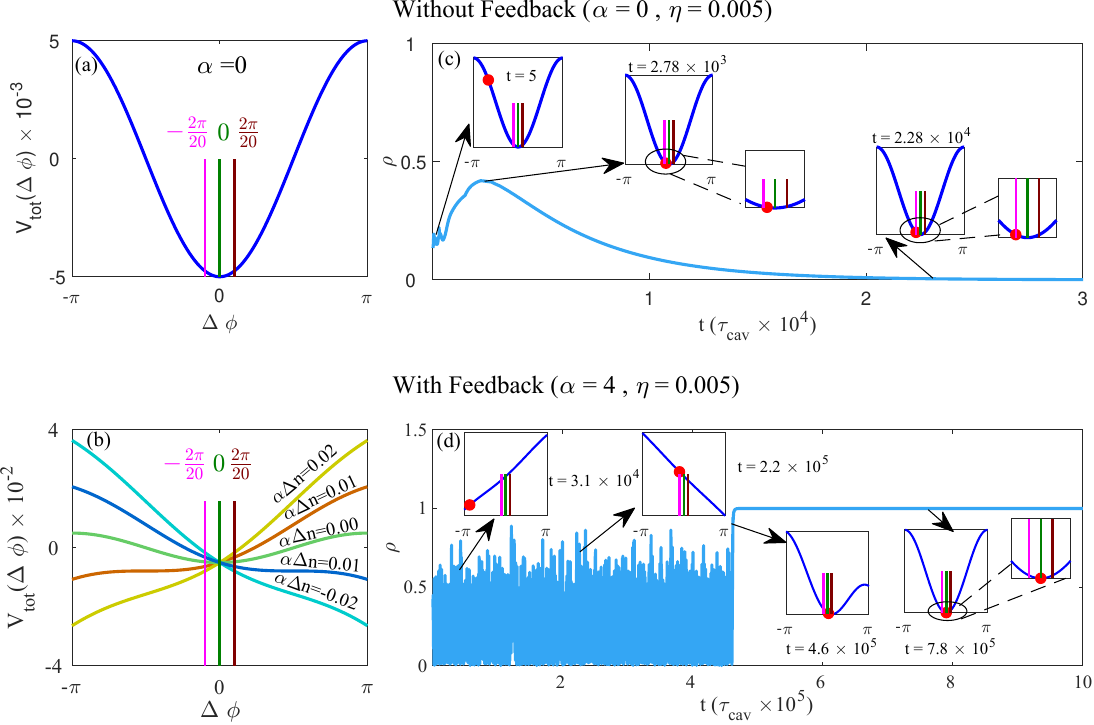}
\caption{Dynamics of the phase difference $\Delta\phi$ for a representative laser pair (lasers 1 and 2) in a 1D ring array of $N = 20$ lasers, described by an effective potential $V_{\mathrm{tot}}(\Delta\phi)$ for $N = 2$. (a) Effective potential without feedback ($\alpha = 0$), shown as the solid blue curve. Vertical lines indicate the phase differences $\Delta\phi = 2\pi/20$ (brown), $0$ (green), and $-2\pi/20$ (pink), corresponding to the $q = +1,~0,~-1$ phase-locked states of the $N = 20$ array, which represent local and global minima in the higher-dimensional phase space of the full array. (b) Effective potential with feedback for different values of $\alpha$. Feedback asymmetrically tilts the potential; the sign of the driving term $\alpha \Delta n$ determines the tilt direction, while its magnitude controls the tilt strength. (c) Time evolution of the order parameter $\rho$ without feedback ($\alpha = 0$). Insets show the instantaneous potential and the effective phase-difference position (red dot) at the indicated times. In the absence of feedback, the potential remains static and the phases become trapped in a local minimum. (d) Time evolution of $\rho$ with feedback ($\alpha = 4$). Strong initial fluctuations in $\rho$ arise from the dynamically evolving, flattened asymmetric potential (insets), enabling the phases to escape from local minima. The later insets show relaxation toward the ground-state ($\rho \rightarrow 1$), where oscillations vanish and the system stabilizes. For panels (b) and (d), the simulation parameters are $\eta=0.005$ and $r=2$.}
\label{Potentail}
\end{figure*}

To gain insight into the physical mechanism of defect suppression induced by intrinsic feedback-driven annealing, we analyze the phase dynamics through an effective potential. While the full $ N=20$ laser ring array evolves in a high-dimensional space, we analyze the evolution of the phase difference between a single pair of neighbouring lasers using an effective potential derived for two coupled lasers ($N=2$). This simplified potential provides a qualitative framework for visualizing how the parameters $\alpha$ and $\eta$ influence phase dynamics and the formation of topological defects. We focus on the evolution of the phase difference \(\Delta \phi = \phi_2 - \phi_1\), as its steady-state value reveals whether the system anneals to the ground state (global minimum) or becomes trapped in a topological defect state (local minimum). Starting from Eq.\,(\ref{Eq2}) and under the approximation \( A_{i+1} / A_{i} \approx 1 \), the dynamics of $\Delta \phi$ is given by:

\begin{equation}
\frac{d(\Delta\phi)}{dt} = -\eta\thinspace \sin(\Delta\phi) + \frac{\alpha}{2}\frac{\Delta n}{n_{\text{th}}}.
\label{eq:4}
\end{equation}
 The first term in Eq.\,(\ref{eq:4}) is the conservative force which can be derived from the static potential $V_{C}(\Delta\phi)=-\eta\thinspace\cos(\Delta\phi)$, i.e. $-\partial V_C(\Delta\phi)/\partial(\Delta\phi)$. For an array with $N>2$, the interaction between each neighboring pair contributes a $-\eta\thinspace$cos$(\Delta \phi_{i,i+1})$ term to the total system potential. This potential is over a high-dimensional phase space (one dimension per laser pair) and contains many local minima, which correspond to topological defect states. While this full landscape cannot be easily visualized, the phase differences for different topological defect states are known (Fig.\,\ref{Fig_1}(b)). The phase differences corresponding to different states can be marked on the one-dimensional potential for $N=2$, providing a simplified visualization model of the final phase-locked states of the larger laser array. Figure\,\ref{Potentail}(a) shows the locations of three stable phase-locked states, marked by vertical lines at phase differences, $2\pi/20$ (brown vertical line), $0$ (green vertical line) and $-2\pi/20$ (pink vertical line) corresponding to topological charges $q=+1$, $q=0$, and $q=-1$, respectively, for a 1D ring array of $N=20$ lasers represented on the effective potential for $N=2$ lasers (solid blue curve).
 
The second term in  Eq.\,(\ref{eq:4}) produces a non-conservative time-dependent driving force, proportional to \(\alpha \, \Delta n\), that couples amplitude fluctuations (via $\Delta n$) to the phase dynamics, $\Delta \phi$. 
To analyse the effect of this driven term on phase dynamics, we define an instantaneous effective potential that combines the static potential with a time-dependent term:
\begin{equation}
V_{\mathrm{tot}}(\Delta \phi, t) = -\eta \cos(\Delta \phi) - 
\left( \frac{\alpha}{2} \frac{\Delta n(t)}{n_{\mathrm{th}}} \right) \Delta \phi,
\label{eq:Vtot}
\end{equation}
where $\Delta n(t)=n_2(t)-n_1(t)$ represents the difference in the carrier density between two lasers at time \(t\).
Figure\,\ref{Potentail}(b) shows the effective potential for different values of $\alpha\thinspace\Delta n$. The potential is symmetric about $\Delta \phi =0$, similar to the static potential (Fig.\,\ref{Potentail}(a)), when $\alpha \Delta n =0$, which corresponds to either no feedback ($\alpha=0$) or stable amplitudes ($\Delta n=0$). For non-zero $\alpha\thinspace\Delta n$, the potential becomes asymmetric, tilted to one side, with the slope increasing in magnitude depending on the value of $\alpha\thinspace\Delta n$, and the sign determining the direction of tilt. This illustrates that the feedback-driven annealing introduces an asymmetry in the potential, which may help to suppress topological defects. 

In the absence of feedback ($\alpha = 0$), no driving term is present; the phase difference evolves solely under the influence of the conservative potential. Figure\,\ref{Potentail}(c) shows one representative evolution of the order parameter $\rho$ for a 1D ring array of $N = 20$ lasers, where the system becomes trapped in a local minimum. The insets display snapshots of the effective potential $V(\Delta\phi)$, with the red point indicating the instantaneous state of the system $(\Delta\phi(t),\, V(\Delta\phi(t)))$, that is, the current phase difference between laser~1 and laser~2 and its corresponding potential energy, at times $t = 5$, $t = 2.78 \times 10^{3}$, and $t = 2.28 \times 10^{4}$ (in units of $\tau_{\mathrm{rav}}$). Initially ($t = 5$), when $\rho$ is small, $\Delta\phi$ is large, placing the system high on the potential slope. With time, $\Delta\phi$ decreases, and the state moves toward the potential minimum, reflected in the increase of $\rho$. However, at $t = 2.78 \times 10^{3}$, the phase difference stuck near $-2\pi/20$ (pink vertical line), corresponding to a local minimum in the high-dimensional potential. With further increase in time, the system remains trapped there and eventually leads to $\rho = 0$, the topological defect state, as shown in the final inset ($t = 2.28 \times10^4$).

On the other hand, with feedback ($\alpha\neq 0$), amplitude and phase fluctuations become strongly correlated, leading to a non-smooth evolution, as previously shown in Figs.\,\ref{Fig_3}(c) and \ref{Fig_3}(d). Figure\,\ref{Potentail}(d) displays a representative evolution of the order parameter $\rho$ under feedback ($\alpha=4$). The four insets show the system’s state in the effective potential landscape at $t = 3.1 \times 10^{4}$, $2.2 \times 10^{5}$, $4.6 \times 10^{5}$, and $7.8 \times 10^{5}$ (in units of $\tau_{\mathrm{rav}}$) (see Appendix\,\ref{sec:appendix E}). During the initial transient, unstable amplitudes make $\Delta n$ large and rapidly fluctuating, allowing the driving term to dominate and the effective potential becomes nearly linear and strongly tilted (inset at $ t=3.1\times10^ {4}$). Because the amplitude fluctuations are fast, the potential slope changes direction frequently (second inset at $t=2.2\times10^{5}$), allowing the system to drift freely and explore the full phase space, reflected in the strong fluctuations of $\rho$. As the system approaches the global minimum, $\Delta n$ decreases, and the conservative part of the potential begins to dominate (third inset at $t=4.6\times10^{5}$), pulling the system toward the ground state configuration ($\Delta\phi = 0$, green vertical line). Eventually, the amplitude stabilizes ($\Delta n=0$), the driving term vanishes, and the system settles into the ground state (fourth inset at $t=7.8\times10^{5}$). Thus, feedback continuously reshapes the potential landscape, effectively shaking the system out of local minima and guiding it towards the lowest-loss in-phase configuration. The efficiency of this process depends on the overall feedback strength, which is governed by the driving term in the effective potential.

From Eq.\,(\ref{eq:Vtot}), the driving term is $\alpha\Delta n$, where $\alpha$ represents the internal amplitude–phase coupling and $\Delta n$ depends on $\eta$, the external coupling that relates phase differences between lasers to amplitude fluctuations. Thus, the effective feedback strength depends on the combined action of $\alpha$ and $\eta$, rather than on either parameter alone. Consequently, increasing the feedback, whether through $\alpha$ or $\eta$, yields similar suppression of topological defects, as confirmed by the phase diagram in Fig.\,\ref{Phase_plot}. This demonstrates that, while both parameters must be non-zero to establish the feedback loop, the suppression strength can be enhanced equivalently by increasing either $\alpha$ or $\eta$.

\section{Topological Defects With Intrinsic Time Scales }
\begin{figure*}[htbp]
\centering
\includegraphics[height = 13.0cm, keepaspectratio = true]{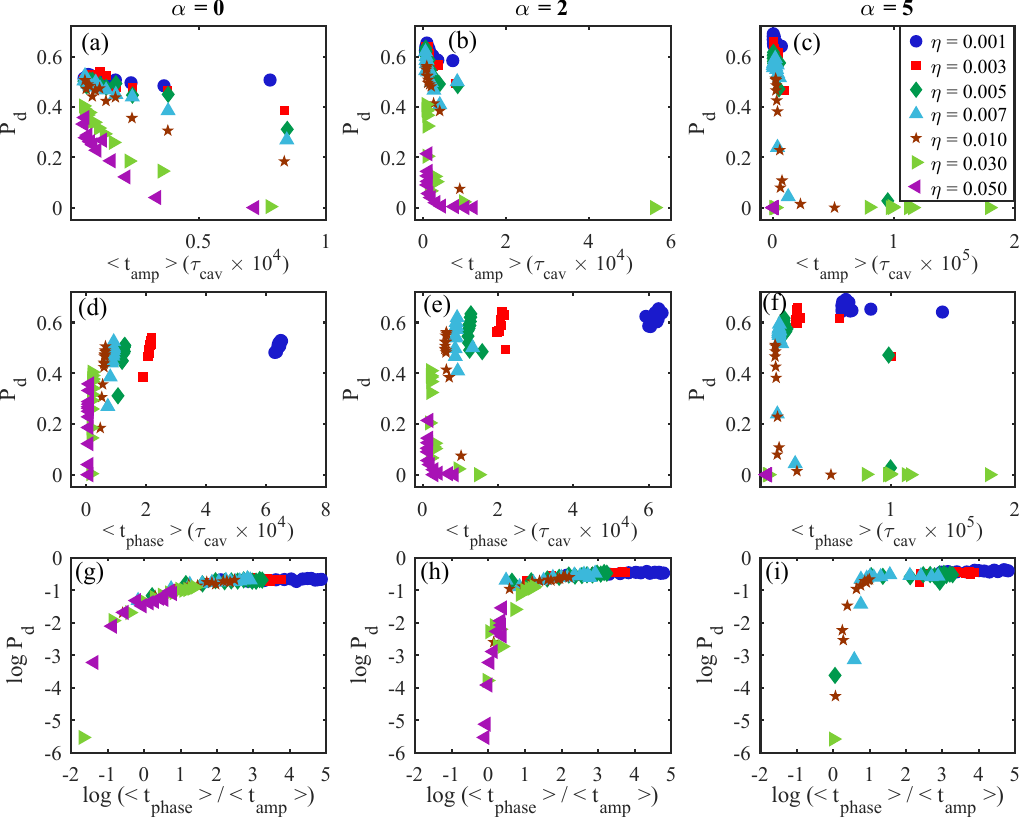}
\caption{Probability of topological defect formation versus amplitude stabilization ($t_{\mathrm{amp}}$) and phase-locking ($t_{\mathrm{phase}}$) time scales. (a)–(c): Defect probability $P_{d}$ as a function of $\langle t_{\mathrm{amp}}\rangle$ for $\alpha=0,~2$ and $5$, respectively. (d)–(f): $\mathrm{P_{d}}$ versus $\langle t_{\mathrm{phase}}\rangle$ for $\alpha=0,~2$ and $5$, respectively. (g)–(i): $\log(\mathrm{P_{d}})$ versus $\log(\langle t_{\mathrm{phase}}\rangle/\langle t_{\mathrm{amp}} \rangle)$ for $\alpha=0,~2$ and $5$, respectively. In each subfigure, for a given value of $\eta$, the different data points correspond to different values of excitation ratio $r$ varied from $2$ to $26$. The quantities $\mathrm{P_{d}}$, $\langle t_{\mathrm{amp}}\rangle$, and $\langle t_{\mathrm{phase}}\rangle$ are obtained by averaging over 2500 realizations of random initial conditions.}
\label{Prob_time_scale}
\end{figure*}
The equivalence between $\alpha$ and $\eta$ in defect suppression suggests that, despite their different physical origins, both parameters influence the laser array dynamics in a similar manner. In annealed spin systems, defect formation is governed by the KZ mechanism, which relates defect formation to the competition between two time scales: the external cooling rate and the internal relaxation time. We therefore expect that defect formation in the laser array is also determined by the competition between two analogous time scales. These are the amplitude stabilization time $t_\mathrm{amp}$, which acts like the external cooling rate, and the phase-locking time $t_\mathrm{phase}$, which corresponds to the internal relaxation time of the system \cite{pal2017}. Both are defined as the times when the amplitude fluctuations $\sigma$ and the order parameter $\rho$ reach within $1\%$ of their steady-state values (see Appendix\,\ref{sec:appendix C}). Since the dynamics of lasers is influenced by several parameters, such as the pump power (excitation ratio), $\alpha$ and $\eta$, we generalize our study by investigating defect formation as a function of the excitation ratio, $\alpha$ and $\eta$. 

Figure\,\ref{Prob_time_scale} shows the probability of topological defect formation $\mathrm{P_{d}}$ as a function of characteristic time scales for a 1D ring array of $N=20$ lasers. The results are presented for different values of $\alpha=0$ (left column), $2$ (middle column), and $5$ (right column), as well as for various values of $\eta$. For each combination of $\alpha$ and $\eta$, the excitation ratio $r$ is varied from $2$ to $26$. Figures\,\ref{Prob_time_scale}(a)–\ref{Prob_time_scale}(c) show $\mathrm{P_{d}}$ as a function of average amplitude stabilization time $\langle t_{\mathrm{amp}} \rangle$; Figures\,\ref{Prob_time_scale}(d)–\ref{Prob_time_scale}(f) show $\mathrm{P_{d}}$ as a function of average phase-locking time $\langle t_{\mathrm{amp}} \rangle$; and Figures\,\ref{Prob_time_scale}(g)–\ref{Prob_time_scale}(i) show $\mathrm{log(P_{d})}$ as a function of ratio between these two time scales ($\log(\langle t_{\mathrm{phase}} \rangle/\langle t_{\mathrm{amp}} \rangle)$). Both $\mathrm{P_{d}}$ and the average time scales $\langle t_\mathrm{amp} \rangle$,  $\langle t_\mathrm{phase} \rangle$, are obtained by averaging over $2500$ random initial conditions for each parameter set. In each subfigure, for a given value of $\eta$, the different data points correspond to different values of $r$.

As evidenced, the probability of topological defects $\mathrm{P_{d}}$ decreases as $t_\mathrm{amp}$ increases for all $\alpha$ (Figs.\,\ref{Prob_time_scale}(a)–\ref{Prob_time_scale}(c)). This is because $t_\mathrm{amp}$ corresponds to the effective cooling rate in the system: a shorter $t_\mathrm{amp}$ represents faster (more abrupt) amplitude relaxation, which promotes defect formation, whereas a longer $t_\mathrm{amp}$ shows slower cooling and suppresses defects, consistent with the general expectation that slower cooling reduces topological defects.
Figs.\,\ref{Prob_time_scale}(d)–\ref{Prob_time_scale}(f) show a subtler dependence of $\mathrm{P_{d}}$ on $t_\mathrm{phase}$. For $\alpha=0$, $\mathrm{P_{d}}$ increases with $t_\mathrm{phase}$ because a longer $t_\mathrm{phase}$ corresponds to slower information exchange across the array (analogous to a longer internal relaxation time in spin systems). When phase information spreads slowly, different regions relax independently, raising the likelihood of topological defects. A similar trend occurs for $\alpha=2$ and $\alpha=5$ at low $\eta$, where feedback is weak. In these regimes, $t_\mathrm{amp}$ and $t_\mathrm{phase}$ act as competing time scales with opposite influences on $\mathrm{P_{d}}$. However, at higher $\eta$ where feedback is strong, the behavior changes significantly. Feedback couples amplitude and phase dynamics, so that amplitudes cannot stabilize unless the system reaches a stable phase-locked state, a signature of feedback-driven annealing, where the effective temperature fluctuates dynamically until the ground-state is approached. This correlation makes $t_\mathrm{phase}$ comparable to $t_\mathrm{amp}$. Consequently, a longer $t_\mathrm{phase}$ now also reduces defect probability, meaning the two time scales influence defects in a similar manner under strong feedback. For each $\alpha$, when $\log(\mathrm{P_{d}})$ is plotted against $\log(\langle t_{\mathrm{phase}} \rangle/\langle t_{\mathrm{amp}} \rangle)$, all the curves corresponding to different $\eta$ collapse into a single curve, as shown in Fig.\,\ref{Prob_time_scale}(g)-\ref{Prob_time_scale}(i). This indicates that, although $t_\mathrm{phase}$ and $t_{\mathrm{amp}}$ individually depend on system parameters such as $r$ and $\eta$, the defect formation is governed by $\langle t_{\mathrm{phase}} \rangle/\langle t_{\mathrm{amp}} \rangle $ only, consistent with the KZ mechanism.

\begin{figure}[htbp]
\centering
\includegraphics[height = 6.0cm, keepaspectratio = true]{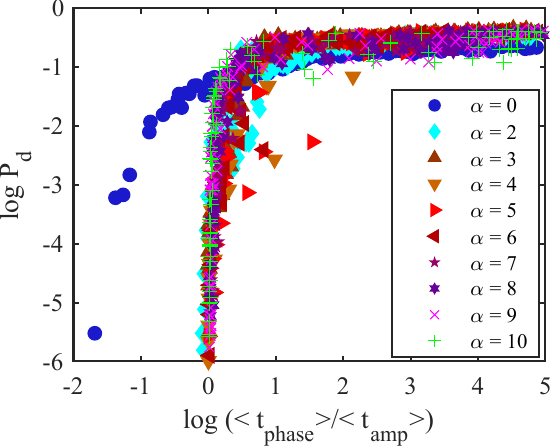}
\caption{ Plot of $\log\mathrm{P_{d}}$ verses log($\langle t_{\mathrm{phase}} \rangle/\langle t_{\mathrm{amp}} \rangle) $ for different values of $\alpha$. For each $\eta$, every point on the curves correspond to different values of excitation ratio $r$. All curves with $\alpha > 0$ collapse onto a single curve, while the uncoupled case ($\alpha = 0$, blue curve) follows a slightly distinct scaling.}
\label{Collapse_all_alpha}
\end{figure}

The collapse of different $\eta$ curves for a fixed $\alpha$ demonstrates that defect formation is governed by the ratio $ \langle t_{\mathrm{phase}} \rangle/\langle t_{\mathrm{amp}} \rangle$. Since the feedback mechanism is identical for different $\alpha$, we expect similar collapse across different $\alpha$ as well. When $\log(\mathrm{P_{d}})$ is plotted versus log($ \langle t_{\mathrm{phase}} \rangle/\langle t_{\mathrm{amp}} \rangle$) for different values of $\alpha$, all curves corresponding to non-zero $\alpha$ collapse into a single curve, while $\alpha=0$  shows slight deviation, as shown in Fig.\,\ref{Collapse_all_alpha}. The different behaviour of $\alpha = 0$ confirms that feedback modifies the dynamics relative to the no-feedback case. An identical curve for all non-zero $\alpha$ reveals a universal feedback-controlled behaviour. This indicates that whenever feedback is present, defect formation is not governed by the individual parameters $\alpha$ and $\eta$, but by the effective feedback strength they produce together. Although different values of $\alpha$ correspond to different semiconductor laser types, the observed universality demonstrates that the feedback–defect mechanism is fundamentally independent of device-specific parameters. This establishes feedback-driven annealing as a robust and tunable approach for suppressing topological defects in coupled laser arrays.

\section{Conclusion}
In conclusion, we have presented an intrinsic feedback-driven annealing mechanism that suppresses the formation of topological defects in coupled class-B semiconductor laser arrays. This mechanism arises from the combined action of internal amplitude-phase coupling $\alpha$ and external coupling $\eta$. Unlike conventional thermal annealing, where the system is externally cooled from high to low temperature, our mechanism relies on self-adjusted amplitude fluctuations: the fluctuation strength automatically increases when the system becomes trapped in a local minimum and decreases as it approaches the ground-state. This dynamic feedback reshapes the effective potential in time, allowing the system to escape local minima and reach the ground-state. We demonstrated that, with an appropriate feedback strength, controlled through either $\alpha$ or $\eta$, the system reaches the ground-state with near 100\% probability, without becoming trapped in local minima. To confirm universality, we analyzed defect formation using two intrinsic timescales, the amplitude-stabilization time $t_{\mathrm{amp}}$ and the phase-locking time $t_\mathrm{phase}$. The defect probability depends only on the ratio $t_\mathrm{phase}/t_{\mathrm{amp}}$, independent of $\alpha$ or $\eta$ individually, consistent with the KZ mechanism. 
Our results establish feedback-driven annealing as a general and robust physical route to achieve defect-free global minimum state in coupled laser arrays. The parameter ranges correspond to realistic semiconductor lasers, making the effect directly testable in VCSEL arrays and degenerate cavity lasers \cite {pal2017,bouchereau2022}. These findings have important implications for optical spin simulators, enabling more accurate and efficient solutions to optimization problems.

\section*{Acknowledgements} \label{sec:acknowledgements}
We acknowledge financial support through the National Quantum Mission (NQM) of the Department of Science and Technology, Government of India. Rajneesh Fulara acknowledges the fellowship support from the University Grants Commission (UGC). We thank Nadia Belabas from C2N, CNRS, France for useful discussions.


\appendix
\section{Steady-state phase-locked solutions of a 1D ring laser array}\label{sec:appendix A}
To identify the steady-state phase-locked solutions of a one-dimensional (1D) ring array of lasers in the absence of internal coupling ($\alpha = 0$), we set $A_i = A$, and detuning $\omega_i = 0$. The dynamical phase equation can be written as:
\begin{equation}
    \frac{d\phi_i}{dt} = \frac{\eta}{2\tau_{\mathrm{cav}}}
    \left[ \sin(\phi_{i+1}-\phi_i) + \sin(\phi_{i-1}-\phi_i) \right].
    \label{eq:phase_cold}
\end{equation}
Allowed steady-state phase-locked solutions will be those for which $\frac{d\phi_i}{dt}=0$. \\
We assume a uniform phase-gradient (constant phase-difference) solution, $\phi_i = m\thinspace i$, where $m$ is the phase increment per site and $i$ is the index of the laser. Substituting $\phi$ into Eq.~\eqref{eq:phase_cold} gives
\begin{align*}
    \frac{d\phi_i}{dt} 
    &= \frac{\eta}{2\tau_{\mathrm{cav}}}\big[ \sin(m) + \sin(-m) \big] \\
    &= \frac{\eta}{2\tau_{\mathrm{cav}}}\big[ \sin(m) - \sin(m) \big] \\
    &= 0.
\end{align*}
Thus, constant-phase-difference solutions produce steady-state phase-locked states.\\
Since the 1D ring consists of $N$ lasers, the physical field must be identical after one complete loop i.e., $ \phi_{i+N} = \phi_i + 2\pi q $, where $q$ is an integer.

\begin{figure}[htbp]
\centering
\includegraphics[scale=0.5]{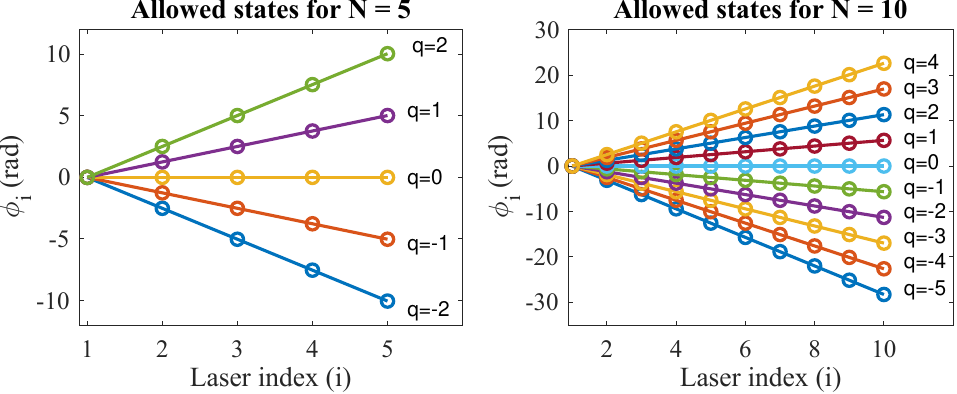}
\caption{Allowed steady-state phase-locked states for 1D ring array of (a) $N=5$ lasers and (b) $N=10$ lasers.}
\label{Allowed_states}
\end{figure}

With $\phi_i = m\thinspace i$ this condition becomes 
\begin{equation}
    mN = 2\pi q \quad \Rightarrow \quad m=\frac{2\pi q}{N}.
\end{equation}
Therefore, the allowed integer values of $q$ that yield $N$ distinct steady-state phase-locked states are
\[
q =
\begin{cases}
-\dfrac{N-1}{2}, \ldots, \dfrac{N-1}{2}, & \text{for odd } N, \\[6pt]
-\dfrac{N}{2}, \ldots, \dfrac{N}{2}-1, & \text{for even } N.
\end{cases}
\]
Integer values of $q$ outside this range correspond to physically equivalent phase configurations within this set. For the 1D ring arrays of $N=5$ lasers and $N=10$ lasers, the number of allowed steady-state phase-locked solutions are shown in Figs.\,\ref{Allowed_states}(a) and \ref{Allowed_states}(b), respectively.
As evident, for $N=5$ there are five steady-state phase-locked solutions corresponding to $q=-2,~-1,~0,~1,~2$, and for $N=10$ there are eleven steady-state phase-locked solutions corresponding to $q=-5,~-4,~-3,~-2,~-1,~0,~1,~2,~3,~4,~5$.

\section{Loss spectrum of different phase-locked solutions}\label{sec:appendix B}
\begin{figure}[htbp]
\centering
\includegraphics[scale=0.50]{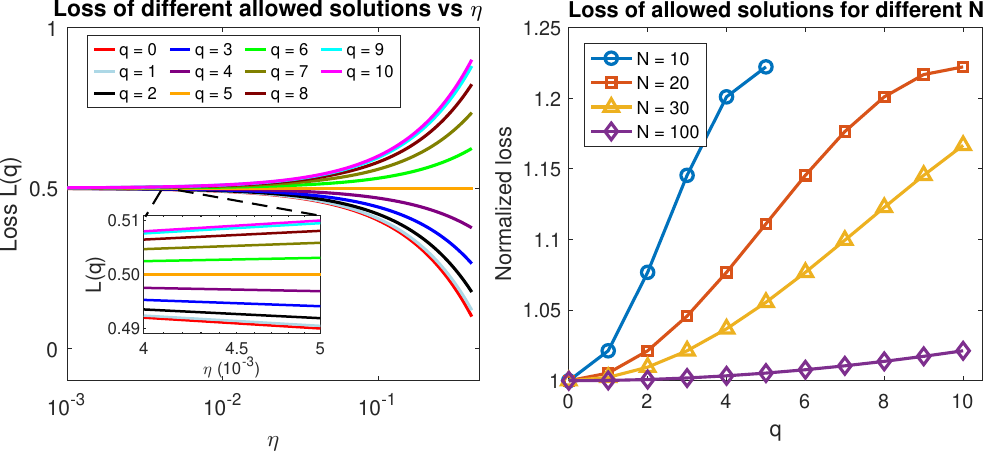}
\caption{(a) In a 1D ring array of $N=20$ lasers, the loss of different allowed phase-locked solutions as a function of external coupling strength ($\eta$), indicating that with increasing $\eta$ the loss difference between different solutions increases. (b) Normalized loss as a function of $q$ for phase-locked solutions, shown for different system sizes ($N$) at a fixed $\eta =0.01$. Note, loss is normalized with respect to threshold pump power of minimum loss solution ($q=0$).}
\label{Loss_modes}
\end{figure}

For a cold cavity (cavity with no gain $n_i/n_{th}=0$), the amplitude equation is given by (in units of cavity round-trip time):
\begin{eqnarray}
    \frac{dA_i}{dt} = -\frac{1}{2}A_i
    + \frac{\eta}{2}\Big[\cos(\phi_{i+1}-\phi_i)A_{i+1} \nonumber\\
    + \cos(\phi_{i-1}-\phi_i)A_{i-1}\Big].
\end{eqnarray}
For a steady-state phase-locked solution $\phi_{i\pm 1}-\phi_i =m=\frac{2\pi q}{N}$ (assuming all lasers have uniform amplitude $A_i=A_{i+1}=A_{i-1}=A$ ), the amplitude evolution of a laser is given by:
\begin{eqnarray}
\frac{dA}{dt}&=& -\frac{1}{2}A+ \frac{\eta}{2}\left[ \cos(m)A + \cos(-m)A \right],\\
    &=& -\frac{1}{2}\left[1 - 2\eta\cos m\right]A.
\end{eqnarray}
Thus, the decay (or growth) rate of \(q_\mathrm{th}\) phase-locked solution is given by:
\begin{equation}
    \lambda_q = -\frac{1}{2}\Big[1 - 2\eta\cos\!\left(\tfrac{2\pi q}{N}\right)\Big], \label{loss-eq}
\end{equation}
which is equivalent to the corresponding cold-cavity loss $L(q)$. The loss of different phase-locked solutions as function of external coupling $\eta$ and for different $N$ is shown in Fig.\,\ref{Loss_modes}.
When gain is included, lasing requires gain to balance loss. The threshold pump power of the \(q_{th}\) phase-locked solution is given by
\begin{equation}
    r_{\text{th}}(q) = L(q).
\end{equation}
The in-phase phase-locked solution ($q=0$) has the lowest threshold:
\begin{equation}
    r_{\text{th}}(0) = \frac{1}{2}(1 - 2\eta).
\end{equation}
\noindent
The losses of phase-locked solutions with different values of $q$ depend on both $\eta$ and $N$ (Eq.\,(\ref{loss-eq}). In particular, the loss difference between distinct phase-locked solutions increases with increasing $\eta$, as shown in Fig.\,\ref{Loss_modes}(a). Furthermore, for a given system size $N$, phase-locked solutions with larger values of $q$ exhibit higher losses (Fig.\,\ref{Loss_modes}(b)). As the system size increases, the loss differences between the various phase-locked solutions become progressively smaller.

\section{Amplitude stabilization time and Phase-locking time}\label{sec:appendix C}
\begin{figure}[htbp]
\centering
\includegraphics[scale=0.54]{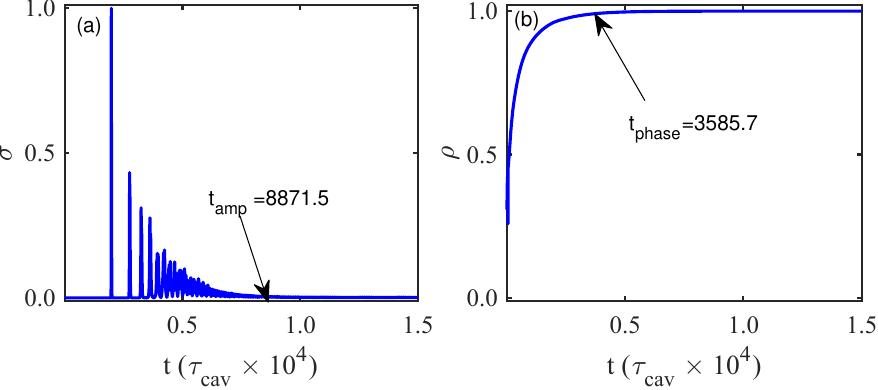}
\caption{In a 1D ring array of N=20 lasers the time evolution of (a) amplitude fluctuations, and (b) order parameter. The amplitude stabilization time ($t_\mathrm{amp}$) and phase-locking time ($t_\mathrm{phase}$) are marked by black arrows in (a) and (b), respectively. The simulation parameters are $r=2$, $\eta=0.01$, and $\alpha=0$.}
\label{Saturation_time}
\end{figure}

Amplitude stabilization time ($t_\mathrm{amp}$) and phase-locking time ($t_\mathrm{phase}$) are calculated from the time dynamics of amplitude fluctuations ($\sigma$) and order parameter ($\rho$), respectively, as shown in the Fig.\,\ref{Saturation_time}. They are defined as the time when the respective parameter reaches within $1\%$ of its final stable value, as marked by arrows in Figs.\,\ref{Saturation_time}(a) and \ref{Saturation_time}(b). The dynamics are mostly stabilized at this time, and there is no major variation in the system dynamics thereafter. 

\section{Dependence of cooling rate on Pump power}\label{sec:appendix D}
\begin{figure}[htbp]
\centering
\includegraphics[scale=0.8]{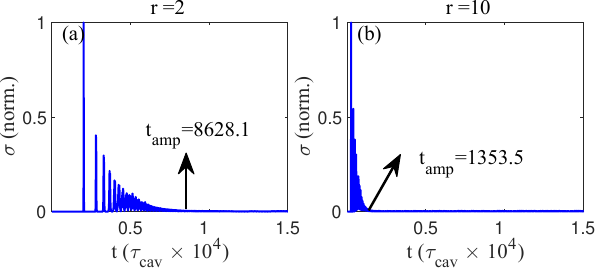}
\caption{In a 1D ring array of N=20 lasers, time evolution of amplitude fluctuations ($\sigma$) for different excitation ratio (a) $r=2$, and (b) $r=10$. Other simulation parameters are $\alpha=0$, $\eta=0.05$.}
\label{Quenching_rate}
\end{figure}

The time-dependent standard deviation of the amplitude ($\sigma$) provides information about the variation of amplitude over time (see Eq.\,(5) in manuscript). The time evolution of $\sigma$ for different excitation ratios, $r=2$ and $10$, is shown in Figs.\,\ref{Quenching_rate}(a) and \ref{Quenching_rate}(b), respectively. As evident, increasing the pump power causes $\sigma$ to approach zero at earlier times (smaller amplitude stabilization time, $t_\mathrm{amp}$ ). Thus, the decay rate of amplitude fluctuations can be controlled through the excitation ratio $r$, indicating that higher values of $r$ lead to shorter amplitude stabilization times. Since the amplitude fluctuations $\sigma$ are analogous to temperature in the annealed spin system, the effective cooling rate of the system can be tuned by adjusting the excitation ratio $r$. 

\section{Variation of Carrier density difference with time}\label{sec:appendix E}
\begin{figure}[htbp]
\centering
\includegraphics[scale=0.70]{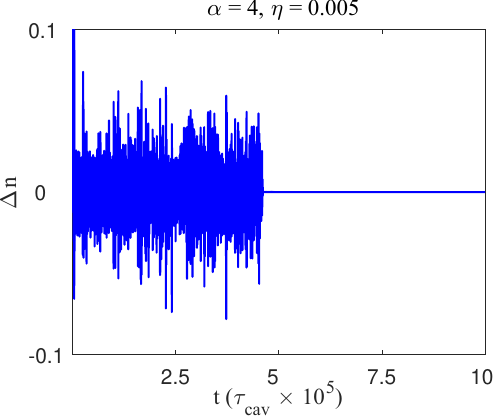}
\caption{Variation of $\Delta n=n_{2}-n_{1}$ with time at $\alpha=4$ and $\eta=0.005$.}
\label{Delta_n_var}
\end{figure}

The carrier density difference $(\Delta n)$ between the lasers plays a crucial role in the dynamics of the class-B semiconductor lasers. In the case of feedback-based annealing, the initial amplitude fluctuations make $\Delta n$ continuously change, as shown in Fig.\,\ref {Delta_n_var}, which results the variations in $\rho$.

The evolution of the phase is determined by the effective potential in Eq.\,(\ref{eq:Vtot}), with $\Delta n(t)=n_2(t)-n_1(t)$ representing the carrier density difference between the two lasers at time \(t\). During the initial transient period, unstable amplitudes make $\Delta n$ large and rapidly fluctuating, allowing the driving term to dominate. As the system approaches the global minimum, $\Delta n$ decreases, and the conservative part (first term on the right side) of the potential begins to dominate, pulling the system toward the in-phase configuration. Finally, the amplitude stabilizes ($\Delta n=0$), the driving term vanishes, and the system settles into the ground-state.

%
\end{document}

%% file: preamble.tex
\usepackage{amsthm}
\usepackage{mathtools}
\usepackage{physics}
\usepackage{xcolor}
\usepackage{graphicx}
\usepackage{float} 
\usepackage{caption}
\graphicspath{{images/}} 
\usepackage[left=20mm,right=20mm,top=35mm,columnsep=15pt]{geometry} 
\usepackage{subcaption}  
\usepackage{wrapfig}  
\usepackage{adjustbox}
\usepackage{placeins}
\usepackage[T1]{fontenc}
\usepackage{lipsum}
\usepackage{csquotes}
\usepackage{ragged2e}